\documentclass[aps,superscriptaddress,showpacs,nofootinbib,eqsecnum,prd,notitlepage,twocolumn]{revtex4}

\usepackage{epsf}  

\usepackage{amssymb}

\usepackage{hyperref}

\usepackage{float}

\usepackage{pgf}

\begin{document}

\title{Degeneracy between CCDM and $\Lambda$CDM cosmologies}
  
\author{Marcelo Vargas dos Santos}
\email{vargas@if.ufrj.br}
\affiliation{Instituto de F\'{\i}sica, Universidade Federal do Rio de Janeiro, C.P. 68528, 
21941-972 Rio de Janeiro, Rio de Janeiro, Brazil}

\author{Ioav Waga}
\email{ioav@if.ufrj.br}
\affiliation{Instituto de F\'{\i}sica, Universidade Federal do Rio de Janeiro, C.P. 68528, 
21941-972 Rio de Janeiro, Rio de Janeiro, Brazil}

\author{Rudnei O. Ramos} \email{rudnei@uerj.br} \affiliation{Departamento de
  F\'{\i}sica Te\'orica, Universidade do Estado do Rio de Janeiro, 20550-013
  Rio de Janeiro, RJ, Brazil}

\begin{abstract}

The creation of
cold dark matter cosmology model is studied beyond the linear
perturbation level. The skewness is explicitly computed and the results
are compared to those from the $\Lambda$CDM model. It is explicitly
shown that both models have the same signature for the skewness and cannot be
distinguished by using this observable. 

\end{abstract}

\maketitle
%%%%%%%%%%%%%%%%%%%%%%%%%%%%%%%%%%%
\section{Introduction}

In a recent work~\cite{rudnei14}, we have investigated the creation of
cold dark matter (CCDM) cosmology as an alternative to explain cosmic
acceleration. The CCDM cosmology~\cite{lima10} is a
phenomenological scenario in which it is assumed that the gravitational field
induces particle creation such that a special choice of the particle
production rate produces a cosmology that, at the background level, is
indistinguishable from the standard $\Lambda$CDM model.   
There have been some recent papers that try to give a more than phenomenological basis
for CCDM particle creation, e.g., Ref.~\cite{jesus}, while a more
detailed study and review of the thermodynamics of cosmologies of
particle creation, including CCDM, has been discussed in
Refs.~\cite{kimura,harko}.
By assuming zero effective sound speed, in Ref.~\cite{rudnei14}, we have compared
CCDM with $\Lambda$CDM showing that these models are observationally
degenerated not only at background but also at the first-order
perturbation level. 

More recently, some of our results have been criticized in
Ref.~\cite{fabris14}. The authors of this paper claim that, since in
CCDM dark matter particles are being continuously created while
baryons are conserved, the ratio between dark matter and baryons
energy densities is not constant and would change with redshift in
these models. According to them, the existence of such variation would
be detectable by current observations breaking the above-mentioned
degeneracy. In fact, they use this argument to rule out CCDM. The key
point here is that to obtain their result, these authors have
considered in their analysis all the amount of dark matter (clustered
or not) in estimating the baryon-dark matter energy densities
ratio. However, what can be measured in gravitational experiments with
clusters of galaxies is only clustered matter, of course. Smoothly
distributed matter or energy at scales of $20-30\;h^{-1}$Mpc, like for
instance, that associated with $\Lambda$, is undetectable in
cosmological tests on these scales. Since in both CCDM and
$\Lambda$CDM, clustered matter (baryons and dark matter) redshifts as
$(1+z)^{3}$ the ratio between dark matter and baryons energy
densities is expected to be constant in both models. Therefore,
contrary to the arguments used  in Ref.~\cite{fabris14}, these models
cannot be observationally distinguished by using, for instance, the
gas mass fraction test~\cite{mantz14}. 

In this paper, we extend the results presented in Ref.~\cite{rudnei14} 
and consider the nonlinear dynamics in the CCDM. 
Here, we explicitly show that the degeneracy between the CCDM and
the $\Lambda$CDM remains at any order in perturbartion theory. 
It is also shown that both models have the same signature for 
the skewness and cannot be distinguished by using this observable.

This paper is organized as follows. In Sec.~\ref{sec1}, we briefly
present the CCDM scenario, emphazising the reason it is not possible to
distinguish  CCDM from $\Lambda$CDM with gravitational experiments
like the gas mass fraction test. In Sec.~\ref{sec2}, we extend our
previous results, obtained in Ref.~\cite{rudnei14}, and consider
nonlinear dynamics in the CCDM models. {}Finally, our conclusions and
final remarks are given in Sec.~\ref{conclusions}.

%%%%%%%%%%%%%%%%%%%%%%%%%%%%%%%%%%%

\section{CCDM model: conserved, created and clustered matter} \label{sec1}

Throughout this work, flat
Friedmann-Robertson-Walker metric is assumed. Since we are mainly interested in
processes that occurred after radiation domination, we neglect
radiation, and, for the sake of simplicity, unless explicitly stated,
we also neglect baryons  considering only the presence (and creation)
of pressureless ($p=0$) dark matter particles. 

The relevant cosmological equations for the CCDM model are ($c=1$)

\begin{equation}
H^{2}=\left( \frac{\stackrel{.}{a}}{a}\right) ^{2}=\frac{8\pi
  G}{3}\rho\,,   
\label{H2}
\end{equation}

\begin{equation}
\frac{\stackrel{..}{a}}{a}=\,\stackrel{\,.}{H}+\,H^{2}=-\frac{4\pi
  G}{3} \left( \rho +3 p_c \right)\,,  
\label{addot}
\end{equation}
where the creation pressure $p_c=-\rho \Gamma/(3H)$ and $\Gamma$ 
is the particle production rate. The fluid equation for
$\rho$ is

\begin{equation}
\stackrel{.}{\rho }+\,3\,H\, \rho  =\rho\, \Gamma\,. 
\label{rho}
\end{equation}
In the CCDM cosmological model, $\Gamma$ is given
by~\cite{rudnei14,lima10}
\begin{equation}
\Gamma=\frac{3\beta
  H_0^2}{H}=3\beta\left(\frac{\rho_{c0}}{\rho}\right)
H\;, \label{gamma}
\end{equation}
where $\beta $ is a $\mathcal{O}(1)$ dimensionless constant,  $H_{0}$
is the current value of the Hubble parameter and $\rho_{c0}\equiv 3
H_0^2/(8\pi G)$ is the critical density at the present time, which, in our
flat-space and simple-fluid approximation, is equal to the value of
dark matter energy density at the present time. With the above choice for
the particle production rate,  Eq.~(\ref{rho}) can be easily
integrated obtaining 
\begin{equation}
\rho=\rho_{c0} \left[ \left( 1-\beta \right)  \left( 1+z\right)^3 +
  \beta \right]\,.
  \label{rho2}
\end{equation}
By substituting Eq.~(\ref{rho2}) in Eq.~(\ref{H2}) we obtain
\begin{equation}
\frac{H^2}{H_{0}^2}=\left(1-\beta \right)  \left( 1+z\right)
^3+\beta\,.
\label{Lcdm}
\end{equation}
The two terms in the right-hand side of Eq.~(\ref{Lcdm}) have a clear
meaning: The first term redshifts exactly as matter, while the second
term, the constant $\beta$, plays the role of the cosmological
constant density parameter at the present time, $\Omega_{\Lambda0}$. The
CCDM model is then able to mimic exactly the $\Lambda$CDM background
expansion history.

In our previous work, Ref.~\cite{rudnei14}, we have split the total
dark matter energy  density $\rho$ in a conserved and created part,
$\rho=\rho_{\rm conserved}+\rho_{\rm created}$, and assumed that the
conserved part of the dark matter energy density is given by
$\rho_{\rm conserved}=\rho_{c0}  \left( 1-\beta \right)  \left(
1+z\right)^3$, while the created one is $\rho_{\rm
  created}=\rho_{c0}\beta$. Although, from the physical point of view,
there is nothing wrong with this choice, it should be remarked that it
is not mandatory since there are other possibilities. To better
understand this statement, notice that from the integration of
Eq.~(\ref{rho}) we get that the conserved energy density part can be
written as 
$\rho_{\rm conserved}=A/a^3$,
while the created part becomes
$\rho_{\rm created}=B/a^3 + \beta  \rho_{c0}$,
where $A$ and $B$ are integration constants. In the special case in
which $\beta=0$,  from Eq.~(\ref{gamma}), we have $\Gamma =0$, and,
since there is no matter creation, it follows from the expression for 
$\rho_{\rm created}$
that we should also have $B=0$. Thus, we can think of $B$ as an
arbitrary  function of $\beta$ such that it vanishes when
$\beta=0$. {}For the sake of simplicity, let us  assume a linear
function, $B=\rho_{c0}\alpha \beta$, where $\alpha$ is a constant. In
this case, it is straightforward to show from the above equations that

\begin{equation}
A= \rho_{c0}\left[ 1-(1+\alpha)\,\beta \right].
\label{A}
\end{equation}

In Ref.~\cite{rudnei14} we have assumed $\alpha=0$. This corresponds
to the special case in which all the energy density of the created
part is constant and all the clustered dark matter  (the part that
redshifts as matter as expected) is conserved. As remarked above, this
choice in not mandatory. Indeed, since the energy density of both
created and conserved parts, should be positive definite, we have the
constraint $0\leq \alpha\leq 1/\beta -1$.
In the case in which $\alpha = 1/\beta -1$, i.e., for the upper
limit for $\alpha$, there is no conserved dark
matter. In other words, in this special case all the  dark matter
(clustered and unclustered) is created during the Universe
evolution. Notice that, since current observations indicate  $\beta
\sim 0.7$ (recalling that $\beta$ in CCDM plays the role of
$\Omega_{\Lambda0}$ in $\Lambda$CDM), the choice $\alpha=1$, made by
the authors of Ref.~\cite{fabris14}, is unphysical since, in this
case, the energy density of the conserved part would be
negative. Thus, it is important to emphasize that in the CCDM scenario
we do not know {\it a priori} which part of the total dark matter particles
has been created and which one is conserved. Only by fixing a value
for $\alpha$ is this choice specified. Although we do not know {\it a priori} which part is created (or conserved), we do know which one
clusters and which one does not. In other words, independent of  the
value of $\alpha$ (the way we split the created and conserved parts
for $\rho$), it is clear from Eq.~(\ref{Lcdm}) that the unclustered
part of the energy density is constant, while the clustered one
redshifts as $a^{-3}$. 

If in addition to dark matter we had considered also baryons, assuming
$B=\rho_{c0}\alpha \beta$ and that baryons are conserved, instead of
Eq.~(\ref{A}) we would get 
$A= \rho_{c0}\left[ 1-(1+\alpha)\,\beta -\Omega_{B0} \right]$,
where $\Omega_{B0}$ is the present value of the baryons density
parameter. The total (baryons+dark matter) energy density will be
$\rho= \rho_{c0}\left[ \Omega_{B0}a^{-3}+\left( 1- \beta
  -\Omega_{B0} \right) a^{-3}+ \beta\right]$.
Therefore, regardless of the value of $\alpha$ again, the ratio
between the baryons energy density ($\rho_B$) and the clustered dark
matter energy density ($\rho_{clust}$) is independent of redshift:
\begin{equation}
\frac{\rho_B}{\rho_{clust}}=\frac{\Omega_{B0}}{1-\beta-\Omega_{B0}}\,.
\end{equation}
As stressed in Ref.~\cite{rudnei14}, it is this ratio, and not
$\rho_B/\rho$, that is estimated, for instance, in x-ray
surveys~\cite{mantz14}. Thus, since it does not depend on redshift, it
will not be possible to distinguish the CCDM scenario from
$\Lambda$CDM by using measurements of the baryon mass fraction in
clusters, as originally suggested in Ref.~\cite{lima12} and considered in
Ref.~\cite{fabris14}.

%%%%%%%%%%%%%%%%%%%%%%%%%%%%%

\section{Dark degeneracy and skewness} \label{sec2}

In Ref.~\cite{rudnei14} we have investigated the growth of linear
perturbations in CCDM models  and compared the neo-Newtonian
\cite{lima} and the general-relativistic frameworks. {}We have shown 
that both approaches
are formally identical only when the effective sound speed ($c_{eff}$)
vanishes  \cite{reis03}.  We also showed, assuming $c_{eff}^2=0$, that CCDM and
$\Lambda$CDM models are degenerate not only at the background level
but at the linear perturbation order as well.  In this section we will
explicitly show that this result is also valid in the nonlinear
regime (nonlinear perturbations have been extensively  extensively examined, for example, in Ref.~\cite{nonlinear}).

Instead of using the neo-Newtonian formulation, as considered in
Ref.~\cite{rudnei14}, in this work, we follow a different, but to some
extent equivalent, approach and start considering Raychaudhuri's
equation for a nonrotating and shearless fluid,

\begin{equation}
\dot{\Theta}+\frac13 \Theta^2= R_{\mu \nu} u^\mu u^\nu\,, 
\label{ray1}
\end{equation}
where $R_{\mu \nu}$ is the Ricci tensor, and in our coordinate system
the fluid 4-velocity is given by $u^\mu=(1,\dot{a}\vec{x}+\vec{v})$,
where $\vec{v}$ is the peculiar velocity. Therefore, $\Theta \equiv
u^\mu_{\,\,\,;\mu}$ can be written as
$\Theta= 3\dot{a}/a+\theta/a$,
where $\theta \equiv \nabla\cdot\vec{v}$. By using Einstein equations,
we can write Eq. (\ref{ray1}) as  

\begin{equation}
\theta'+\frac{ \theta}{a}  + \frac{\theta^2}{3 H}=-\frac{4 \pi G}{H}
\left(\delta\rho+3\delta P\right)\,,
\label{ray2}
\end{equation}
where the prime denotes differentiation with respect
to the scale factor $a$ and, as usual, we decompose the dynamical
variables into their background and inhomogeneus parts; i.e., we
write $ \rho = \tilde{\rho} + \delta\rho = \tilde{\rho}(1+\delta)$
and $P    = \tilde{P} + \delta P$. Here, the tilde denotes background 
quantities. 

By using the conservation equation
$
\dot{\rho}+\left(\rho + P\right) \Theta= 0 , \label{drho}
$
we obtain that the density contrast $\delta$ satisfies the
differential equation
\begin{equation}
 \delta ' +\frac{3}{a} (c_{eff}^2-w)
 \delta+\left[1+w+\delta(1+c_{eff}^2)\right]\frac{\theta}{H a^2} =
 0, \label{del}
\end{equation}
where $w=\tilde P /\tilde \rho$ and $c_{eff}^2=\delta P / \delta
\rho$.

Assuming $c_{eff}^2=0$, differentiating  (\ref{del}) with respect to
the scale factor and using (\ref{ray2}), after some algebra, we obtain
the following differential equation for the density contrast:

\begin{eqnarray}
\lefteqn{ a^2 \delta '' +a \delta ' \left( \frac{3- 9 w }{2}- 
\frac{ a w'}{1+w+\delta}\right) - \frac{4 a^2 {\delta '}^2 }
{3 (1+w+\delta)} }
\nonumber \\
&&\!\!\!\!\!\!\!+ \frac{5 a \delta \delta' w}{1+w+\delta} + 
\frac{3\delta}{2} \left( 3w^2-2w-1 -2 a w' +\frac{2 a w
      w'}{1+w+\delta}\right)
\nonumber \\
&&\!\!\!\!\!\!\!  - 
3\delta ^2 \left(\frac{w^2}{1+w+\delta}+\frac12\right) =0 .
\label{delta2na}
\end{eqnarray}
The same differential equation for the density contrast as above was
obtained in  Ref.~\cite{reis04} by using the neo-Newtonian
formulation (note that in Ref.~\cite{reis04} the derivatives were taken
with respect to the conformal time).
As discussed in Ref.~\cite{rudnei14}, assuming $c_{eff}^2\neq 0$ introduces a
scale dependence  of the perturbations even at linear order. This
scale dependence can cause strong  oscillations if $c_{eff}^2 > 0$, or
exponential growth if $c_{eff}^2 < 0$. Thus, only  models with
$|c_{eff}^2|\ll 1$ are acceptable at linear scales.

Since we are interested in studying the weakly nonlinear regime of
structure formation and to compute higher-order moments of the density
distribution it is useful to  expand $\delta$ as \cite{bernardeau92},

\begin{equation}
\label{expansion}
\delta=\sum_{i=1}^{\infty}\delta_i=\sum_{i=1}^{\infty}\frac{D_i(a)}
           {i!}\delta_{0}^{i}\,, 
\end{equation}
where $\delta_{0}$ is a small perturbation. 
{}For special models, like CCDM, in which the adiabatic sound speed
is equal to zero and recalling that
$c_s^2={\tilde{P}'}/{\tilde{\rho}'}=w- a {w}'/[3(1+w)]$, we obtain that
$D_1$ satisfies

\begin{equation}
\!\!\!D_1'' + \frac{3}{2a}\left(1-5w \right) D_1' + \frac{3}{2a^2} 
\left( 3 w^2-8w -1 \right)D_1 = 0. 
\label{d1b}
\end{equation}
By using that in CCDM 
$w(a)=-\beta/[\beta+(1-\beta)a^{-3}]$,
Eq.~(\ref{d1b}) can be integrated and, in terms of the
hypergeometric functions,  $_2F_1(a,b;c;x)$, the growing mode can be
expressed as~\cite{rudnei14}
\begin{equation}
D_1(a,\beta) = \frac{a}{1+\frac{a^3\beta}{1-\beta}}\;  {}_2 F_1
\left(\frac13, 1; \frac{11}{6}; -\frac{a^3\beta}{1-\beta}\right).
\label{D1}
\end{equation}

As remarked in our
previous work \cite{rudnei14}, at first order in CCDM, as we increase
$\beta$, there is a density contrast suppression as compared to
$\Lambda$CDM.
The suppression factor is given by
$1/[1+ \beta a^3/(1-\beta)]=1+w(a)$
and, as it will be demonstrated below, remains in any order of
perturbation theory. As discussed in Ref.~\cite{rudnei14} the presence of
this suppression is related to the fact that in CCDM dark matter
clusters in the same manner as it does in $\Lambda$CDM. The suppression appears when the constant and nonclustered part of CCDM energy density starts to become non-negligible, and, as a consequence,
the equation of state parameter deviates from zero. However, as we
have already pointed out, in CCDM when considering tests that involve
the growth factor like, for instance, the redshift-space-distortion
$f(z) \sigma_8(z)$ test, what should be considered in the calculation
of these quantities is only the clustered part of $\rho$ (we direct the
interested reader to Ref.~\cite{rudnei14}, where 
this issue has been more closely discussed).

Let us now illustrate the validity of the above result also in
second order of perturbation theory. The second-order solution is
obtained by using in Eq.~(\ref{delta2na}) that $\delta=D_1 \delta_0 + 
D_2 \delta^2_0/2$ and $1/(1+w+\delta) = 1/(1+w)- 1/(1+w)^2 
\delta +{\cal O}(\delta^2)$.
We then find that the second-order factor in the expansion
(\ref{expansion}), when keeping only second-order terms in $\delta_0$, satisfies the differential  equation

\begin{eqnarray}
\lefteqn{\!\!\!\!\!\!\!\!\!\!\!\!\!\!\!\!\!\! {D_2}'' \!+ 
\frac{ 3 {D_2}'} {2 a} \left( 1- 5 w\right) \! +
  \frac{3 D_2 }{2a^ 2}  \left( 3w^2-8w-1\right) - \frac{8 {D_1'}^2}{3(1+w)} }
\nonumber \\
&& +  \frac{16 D_1
    {D_1}' w}{a(1+w)} \! - \! \frac{ 3{D_1} ^2}{a^ 2}
  \left(\frac{8 w^2}{1+w}+1\right)=0  .
\label{D2}
\end{eqnarray}
Analogously, higher-order terms are obtained recursively by using the
solutions of the differential equations for the lower-order ones. 
The solution for $D_2$ in CCDM is obtained by numerically
integrating Eq.~(\ref{D2}), using the solution $D_1$ in CCDM given by
Eq.~(\ref{D1}), and assuming initial conditions such that $D_1$ and $D_2$
at high redshift (small values of the scale factor) behave like in an
Einstein-de Sitter model ($D_1 \propto a$, $D_2 \propto a^2$ and such that the skewness assumes the value $S_3 = 34/7$ at that time). 
The solution $D_2$ for $\Lambda$CDM is obtained
analogously by numerical integration of the differential equation,
equivalent to Eq.~(\ref{D2}), valid in the $\Lambda$CDM case (see also, e.g.,
Ref.~\cite{Multamaki}).

\begin{figure}[htb]
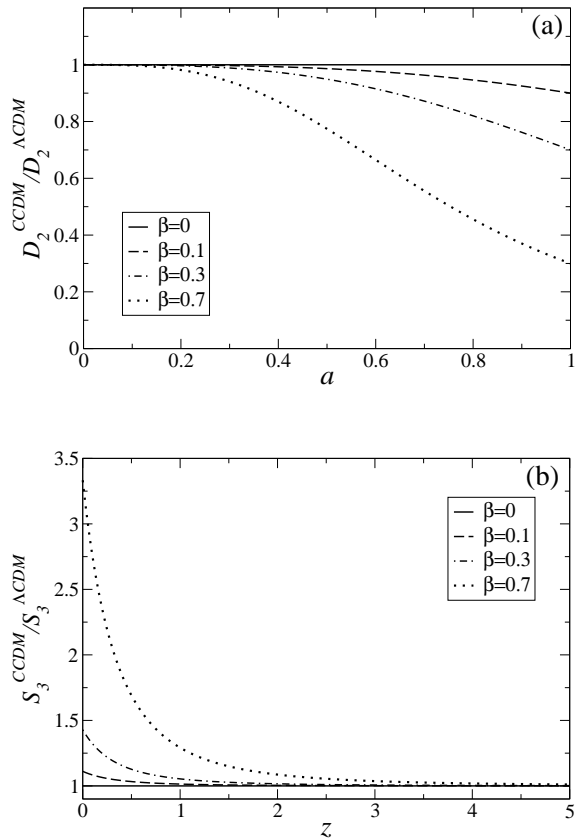

\begin{center}
\begin{tabular}{ccc}
\includegraphics[width=7.5cm]{darkdeg-fig1.eps} \\ \\ \\
\includegraphics[width=7.5cm]{darkdeg-fig2.eps} 
\end{tabular}
\end{center}
\caption{(a) The ratio of growth function at second order $D_2$,
as a function of the scale factor, and (b) the ratio of skewness $S_3$,
as a function of the redshift,
between the CCDM and $\Lambda$CDM models
and for different values of $\beta$. In both cases, the
total density $\rho$ is included, Eq. (\ref{rho2}), for the CCDM. When including only
the clustering part (see the text), $w=0$,  $\rho^{CCDM} \to \rho_m^{\Lambda CDM}$,
and both CCDM and $\Lambda$CDM
become fully degenerate.}
\label{fig} 
\end{figure}

In  {}Fig.~\ref{fig}(a), we show the ratio for $D_2$ between CCDM
and $\Lambda$CDM. It is clear again the presence of
suppression in CCDM as compared to $\Lambda$CDM  as we increase
$\beta$.  The suppression is found again to be such that $D_2^{CCDM} =
[1+w(a)]D^{\Lambda CDM}_2$.  This result holds for any order in
perturbation theory and it can be proved as follows.  {}First note
from Eq.~(\ref{rho2}) that $\rho^{CCDM}(a) = \rho_{c0}
\left[(1-\beta)/a^3 + \beta \right]$, while $\rho_m^{\Lambda CDM} =
\rho_{c0} \Omega_{m0} /a^3$, with $\Omega_{m0}= 1-\beta$.  Thus, at
any order of perturbation theory it holds that $\delta \rho = \delta
\rho_m$ for a constant $\beta$, and then, we also find that
$\rho_m^{\Lambda CDM} = [1+w(a)] \rho^{CCDM}$. Hence, the density
contrast in each case is then simply related by $\delta^{CCDM} =
[1+w(a)] \delta^{\Lambda CDM}$. Using this result back in the
definition  (\ref{expansion}), it immediately allows us to conclude
that at any $n$th-order in  perturbation theory $D_n^{CCDM} =
[1+w(a)]D^{\Lambda CDM}_n$. This concludes our proof.

Returning again to the second-order result, $D_2$, and assuming
Gaussian initial conditions, we can associate $D_2$ with the
emergence of non-Gaussian features in the matter density
field. Indeed, $D_2$ is related to the skewness of the cosmic
field~\cite{peebles,fry}. The skewness is defined by 

\begin{equation}
S_3=3D_2/D^{2}_{1}.
\end{equation} 

We show in Fig.~\ref{fig}(b) the ratio for the skewness between the CCDM 
the $\Lambda$CDM models. The numerical result, in fact, can be expressed
analytically as
$S_3^{CCDM}(a)= 1/[1+w(a)] S^{\Lambda CDM}_3 (a)$. In particular, $S_3(z=0)=
1/(1-\beta) S^{\Lambda CDM}_3 (z=0)$. For $\Lambda$CDM, the skewness is 
nearly constant, with $S^{\Lambda CDM}_3 \approx 4.86$, and is weakly 
sensitive to $\Omega_{m0}$.  One
may be tempted to interpret that this difference is an indication that
CCDM models with $\beta \simeq 0.7$ ($\Omega_{clust0}\simeq 0.3$) are
inconsistent with large-scale skewness measurements
\cite{skewobs}. However, care should be taken when analyzing this
issue for two reasons. First, in our discussion, baryons were
neglected, and measurements of skewness from large-scale galaxy
distribution are based on counting luminous objects, not the dark
component. Our analysis can easily be generalized  to include a small
amount of baryons as observed ($\Omega_{b0}\approx0.04$). It can be
shown that in CCDM    the baryonic skewness does not change much with
redshift, it is nearly constant such that  $S_{3b} \simeq 4.86$ as
expected for dark matter and/or baryons in $\Lambda$CDM. Therefore,
skewness from large-scale galaxy distribution will not be able to
break the degeneracy. Second, one may still argue that lensing (convergence)
skewness~\cite{bernardeau02}, could break the degeneracy between CCDM
and $\Lambda$CDM. However, lensing skewness is sensitive to clustered
matter (baryons and dark matter) and since dark matter in CCDM
clusters in the very same manner as it does in $\Lambda$CDM,
measurements of lensing skewness will also not be able to break the
degeneracy between the models.

\section{Discussion and Conclusions} 
\label{conclusions}

A common assumption in cosmology is that we live in a Universe with a
dark sector composed by two separately conserved components:
clustering dark matter, responsible for large-scale structure
formation, and a nonclustering dark energy, responsible for the cosmic
acceleration. Cosmologies, like $\Lambda$CDM, with the dark sector
defined in this manner can fit well the observations. However, this
kind of division of the dark sector is not unique and is, in a certain
sense, arbitrary. As a matter of fact, we have physically different
cosmologies, based on distinct assumptions and/or a separation of the
dark sector, that cannot be observationally distinguished from the two
dark components cosmologies. Different aspects of this degeneracy in
the dark sector, and how cosmological models can be observationally
distinguished, have been considered in the literature by several
authors (see, for instance,
Refs.~\cite{kunz09,wasserman02}). Dark degeneracy is what this property
has been called~\cite{kunz09}. 

Perhaps, the first controversy we find in the literature regarding the
above-mentioned difficulty is related to the $\Lambda$CDM limit of the
so-called generalized Chaplygin gas  (GCG) cosmological
model~\cite{avelino03,fabris04}. Indeed, in Ref.~\cite{avelino03} (see
also Ref.~\cite{sandvick04}), it has been shown that gravity alone
cannot distinguish the $\alpha=0$ quartessence GCG model from
$\Lambda$CDM. 
The basis of the results we have presented in this paper is the same.
In other words, under certain assumptions
like, for instance, zero effective sound speed, we cannot distinguish with cosmological observations CCDM from $\Lambda$CDM (and, of course,
also the $\alpha=0$ GCG model). 

By assuming zero effective sound speed, we have shown in
Ref.~\cite{rudnei14} that  CCDM and $\Lambda$CDM are observationally
degenerated at both the background and at first-order perturbation
levels. In this paper, we have extended those results considering
nonlinear dynamics in these models. 
In particular we have shown that
they have the same signature for the skewness and explained why they
cannot be distinguished by using this observable.

\acknowledgments

The authors would like to thank Saulo Carneiro, Julio Cesar Fabris,
Jose Ademir Sales de Lima, and Jose Antonio de Freitas Pacheco for
useful discussions.  R.O.R and I.W. are partially supported by
Conselho Nacional de Desenvolvimento Cient\'{\i}fico e Tecnol\'ogico
(CNPq). M.V.S. thanks Funda\c{c}\~ao Carlos Chagas Filho de Amparo \`a Pesquisa do Estado do
Rio de Janeiro (FAPERJ) for support.
R.O.R is also partially supported by a research grant from
FAPERJ.


\begin{thebibliography}{99}

\bibitem{rudnei14} R.~O.~Ramos, M.~V.~dos Santos, and I.~Waga, 
%``Matter creation and cosmic acceleration,'' 
  Phys.\ Rev.\ D {\bf 89}, 083524 (2014).
%  [arXiv:1404.2604 [astro-ph.CO]].
  %%CITATION = ARXIV:1404.2604;%%

\bibitem{lima10} J.~A.~S.~Lima, J.~F.~Jesus, and F.~A.~Oliveira,
  %``CDM Accelerating Cosmology as an Alternative to LCDM model,''
  JCAP {\bf 11},  027 (2010).
%  [arXiv:0911.5727 [astro-ph.CO]].
  %%CITATION = ARXIV:0911.5727;%%

\bibitem{jesus}J.~F.~Jesus and S.~H.~Pereira,
  %``CCDM model from quantum particle creation: constraints on dark matter mass,''
  J. Cosmol. Astropart. Phys. 07 ({\bf 2014}) {\bf 1407}, 040 (2014);
%  [arXiv:1403.3679 [astro-ph.CO]].
  %%CITATION = ARXIV:1403.3679;%%;
J. A. S. Lima and I. Baranov,  Phys. Rev. D {\bf 89},  043515 (2014). 

\bibitem{kimura} N.~Komatsu and S.~Kimura,
  %``Entropic cosmology in a dissipative universe,''
  arXiv:1408.4836;
  %%CITATION = ARXIV:1408.4836;%%
N.~Komatsu and S.~Kimura,
  %``Evolution of the universe in entropic cosmologies via different formulations,''
  Phys.\ Rev.\ D {\bf 89}, 123501 (2014).
%  [arXiv:1402.3755 [astro-ph.CO]].
  %%CITATION = ARXIV:1402.3755;%%

\bibitem{harko}T.~Harko,
  %``Thermodynamic interpretation of the generalized gravity models with geometry - matter coupling,''
  Phys.\ Rev.\ D {\bf 90}, 044067 (2014).
%  [arXiv:1408.3465 [gr-qc]].
  %%CITATION = ARXIV:1408.3465;%%

\bibitem{fabris14}J.~C.~Fabris, J.~A.~de Freitas Pacheco and O.~F.~Piattella,
  %``Is the continuous matter creation cosmology an alternative to $\Lambda$CDM?,''
  J. Cosmol. Astropart. Phys. 06 (2014) 038 (2014).
%  [arXiv:1405.6659 [astro-ph.CO]].
  %%CITATION = ARXIV:1405.6659;%%

\bibitem{mantz14} A.~B. Mantz, S.~W. Allen, R.~G. Morris, D.~A. Rapetti, D.~E. Applegate, P.~L. Kelly, A. von der Linden and R.~W. Schmidt, Mon. Not. R. Astron. Soc. \textbf{440}, 2077 (2014) ; S.~W. Allen, A.B. Mantz, R. G. Morris, D. E. Applegate, P. L. Kelly, A. von der Linden, D. A. Rapetti, and R. W. Schmidt,  arXiv:1307.8152.

\bibitem{lima12}J.~A.~S.~Lima, S.~Basilakos and F.~E.~M.~Costa,
  %``New Cosmic Accelerating Scenario without Dark Energy,''
  Phys.\ Rev.\ D {\bf 86},  103534 (2012).
%  [arXiv:1205.0868 [astro-ph.CO]].
  %%CITATION = ARXIV:1205.0868;%%

\bibitem{lima}  J.~A.~S. Lima, V. Zanchin and R.~H. Brandenberger,
  Mon.\ Not.\ R.\ Astron.\ Soc.\  {\bf 291},  L1 (1997).

\bibitem{reis03}  R.~R.~R.~Reis,
  %``Domain of validity of the evolution of perturbations in Newtonian cosmology with pressure,''
  Phys.\ Rev.\ D {\bf 67},  087301 (2003)
   [Erratum-ibid.\ D {\bf 68},  089901 (2003)].
  %%CITATION = PHRVA,D67,087301;%%

\bibitem{nonlinear}H. Noh and J. c. Hwang, Phys. Rev. D {\bf 69}, 104011 (2004); N.
Bartolo, E. Komatsu, S. Matarrese, and A. Riotto, Phys. Rep. {\bf 402}, 103
(2004); D. H. Lyth, K. A. Malik, and M. Sasaki, J. Cosmol. Astropart.
Phys. {\bf 05} (2005) 004; D. Langlois and F. Vernizzi, Phys. Rev. Lett. {\bf 95},
091303 (2005).

\bibitem{reis04} R.~R.~R. Reis, M. Makler, and I. Waga, Phys. Rev. D {\bf 69},  101301(R) (2004).

\bibitem{bernardeau92} F. Bernardeau, Astrophys. J. \textbf{392}, 1
(1992); \textbf{433}, 1
(1994);
P. Fosalba and E. Gazta\~{n}aga,  Mon. Not. R. Astron. Soc. \textbf{301}, 503
(1998).

\bibitem{Multamaki} 
  T.~Multamaki, E.~Gaztanaga, and M.~Manera,
  %``Large scale structure in nonstandard cosmologies,''
  Mon.\ Not.\ R.\ Astron.\ Soc.\  {\bf 344}, 761 (2003).
  %%CITATION = ASTRO-PH/0303526;%%

\bibitem{peebles} P.~J.~E. Peebles, \emph{The Large Structure of the Universe}, (Princeton University Press, Princeton, NJ, 1980).

\bibitem{fry} J. Fry, Astrophys.J. \textbf{279}, 499 (1984).

\bibitem{skewobs} E. Gazta\~{n}aga and J.~A. Frieman,  Astrophys. J. \textbf{437}, L13 (1994); F. Hoyle, I.
Szapudi, and C.~M. Baugh,  Mon. Not. R. Astron. Soc. \textbf{317},
L51 (2000); I. Szapudi, M. Postman, T.~R. Lauer, and W. Oegerle,
Astrophys. J. \textbf{548}, 114 (2001); I. Szapudi {\it et. al.},
Astrophys. J. \textbf{570}, 75 (2002); for review see F.
Bernardeau, S. Colombi, E. Gazta\~{n}aga, and R. Scoccimarro,
Phys. Rep. \textbf{367}, 1 (2002).

\bibitem{bernardeau02} F. Bernardeau, L. van Waerbeke , and Y. Mellier,
Astron. Astrophys. \textbf{322}, 1 (1997).

\bibitem{kunz09} M. Kunz, Phys. Rev.D \textbf{80}, 123001 (2009); M. Kunz, J. Phys. Conf. Ser. 110, 062014 (2008); 

\bibitem{wasserman02} I. Wasserman, Phys. Rev. D \textbf{66}, 123511 (2002); C. Rubano and P. Scudellaro, General Relativ. Gravit. \textbf{34}, 1931 (2002); A. Aviles and J.~L. Cervantes-Cota, Phys. Rev. D 
\textbf{84}, 083515 (2011); S. Nesseris, Phys. Rev. D \textbf{88}, 123003 (2013); S. Carneiro and H.~A. Borges, J. Cosmol. Astropart. Phys. {\bf 06}, 010 (2014).

\bibitem{avelino03} P.~P. Avelino, L.~M.~G. Be\c{c}a, J.~P.~M. de Carvalho, and C.~J.~A.~P. Martins, J. Cosmol. Astropart. Phys. {\bf 09}, 002 (2003).

\bibitem{fabris04} J.~C. Fabris, S. Gon\c{c}alves, and R.S. Ribeiro, General Relativ. Gravit. \textbf{36}, 211 (2004).

\bibitem{sandvick04} H. Sandvick, M. Tegmark, M. Zaldarriaga and I. Waga,  Phys. Rev. D \textbf{69}, 123524 (2004).

\end{thebibliography}
\end{document}